\documentclass[fleqn,twoside]{article}
\usepackage{espcrc2}
\newcommand{\beq}{\begin{equation}}
\newcommand{\eeq}{\end{equation}}
\newcommand{\ba}{\begin{eqnarray}}
\newcommand{\ea}{\end{eqnarray}}

\title{On the Role of Boundary Terms in the AdS/CFT Correspondence}
\author{Pablo Minces\address{Instituto de F\'{\i}sica Te\'orica, 
Universidade Estadual
Paulista,\\ Rua Pamplona 145, 01405-900, S\~ao Paulo, SP, Brasil}%
\thanks{I would like to thank Prof. V. O. Rivelles for the 
collaborations and discussions on this topic. This work was supported 
by FAPESP grant 01/05770-1.}}

\begin{document}

\begin{abstract}
We consider a scalar field theory on AdS, and show that the usual AdS/CFT 
prescription is unable to map to the boundary a part of the information 
arising from the quantization in the bulk. We propose a solution to this 
problem by defining the energy of the theory in the bulk through the 
Noether current corresponding to time displacements, and, in addition, by 
introducing a proper generalized AdS/CFT prescription. We also show how 
this extended formulation could be used to consistently describe 
double-trace interactions in the boundary. The formalism is illustrated 
by focusing on the non-minimally coupled case using Dirichlet boundary 
conditions.

\end{abstract}

\maketitle

In recent years, and after the work by Maldacena \cite{maldacena} 
conjecturing the 
existence of a duality between a supergravity theory on Anti-de Sitter 
(AdS) space and a conformal field theory (CFT) living at its boundary, a 
large amount of work has been devoted to get a deeper understanding of the 
so-called AdS/CFT correspondence. In particular, it was given a 
prescription which explicitly maps one theory into the other 
\cite{witten}\cite{gubser}. It reads
\ba
Z _{AdS}[\phi _{0}] &=& \int _{\phi _{0}}{\cal D}\phi\
\exp\left(-I[\phi]\right)\equiv Z _{CFT}[\phi _{0}] 
\nonumber\\
&=& \left<\exp\left(\int
_{\partial\Omega}d^{d}x{\cal O}\phi _{0}\right)\right>\; ,
\label{9003}
\ea
where $d+1$ is the dimension of the AdS space, and $\phi_{o}$ is the 
boundary value of the bulk field $\phi$ which couples to the boundary CFT 
operator ${\cal O}$.

Throughout this paper, we will focus on the formulation of a massive, 
non-minimally coupled scalar field theory on AdS. We will show the 
existence of some unsolved difficulties and suggest possible 
solutions to them. We begin by considering the action
\ba
I_{0}&=&-\frac{1}{2}\;\int d^{d+1} x \;\sqrt{g}\;
[
g^{\mu\nu}\partial_{\mu}\phi\;\partial_{\nu}\phi\nonumber\\ 
&&\qquad\qquad\qquad +
\;\left( m^{2}+\;\varrho R\right)\phi^{2}] \; ,
\label{9016}
\ea
where $R$ is the Ricci scalar corresponding to $AdS_{d+1}$, and it is a 
constant. Here $\varrho$ is an arbitrary coupling coefficient.

In particular, we will be concerned with old and well-established results
regarding the quantization of the above theory in global
coordinates. It
is known \cite{freedman}\cite{freedman8}\cite{mezincescu} that there exist
two possible quantizations involving two different normalizable modes,
namely, regular and irregular ones. They behave close to the boundary as
\beq
\phi_{R}\sim{\hat\epsilon}^{\Delta_{+}(\varrho)}\; ,\quad
\phi_{I}\sim {\hat\epsilon}^{\Delta_{-}(\varrho)}\; ,
\label{9006'}
\eeq
where $\phi_{R}$, $\phi_{I}$ correspond to regular and irregular
modes respectively, ${\hat\epsilon}$ is a measure of the distance to the
boundary which is considered to be small, and
\beq
\Delta_{\pm}(\varrho)=\frac{d}{2}\;\pm\;\nu(\varrho)\; ,
\label{9006''}
\eeq
\beq
\nu(\varrho)=\sqrt{\frac{d^2}{4}\;+\;m^{2}+\;\varrho R}\; .
\label{9006'''}
\eeq
One important result in \cite{freedman}\cite{freedman8} is
that irregular modes are normalizable only for
\beq
0\leq \nu(\varrho) < 1\; .
\label{12000}
\eeq
In addition, the energy for irregular modes is conserved, positive and 
finite only when the following constraint is satisfied
\beq
\varrho = \frac{1}{2}\;\frac{\Delta_{-}(\varrho)}
{1+2\Delta_{-}(\varrho)}\; .
\label{9041}
\eeq

Since there are two possible quantizations in the bulk, we expect to 
find two different boundary CFT's, corresponding to the conformal 
dimensions $\Delta_{+}(\varrho)$ and $\Delta_{-}(\varrho)$. However, 
the 
prescription Eq.(\ref{9003}) reproduces only 
$\Delta_{+}(\varrho)$. Note, also, that $\Delta_{+}(\varrho)$ is bounded 
from 
below by $\frac{d}{2}\;$, which is more stringent than the unitarity 
bound $\frac{d-2}{2}\;$. In order to also account for the missing 
conformal dimension $\Delta_{-}(\varrho)$, the proposal in 
\cite{witten2} 
was that its generating functional could be obtained by performing a 
Legendre transform on the original one corresponding to the conformal 
dimension $\Delta_{+}(\varrho)$. Thus, starting from the generating 
functional in momentum space corresponding to $\Delta_{+}(\varrho)$ 
\cite{witten}\cite{freedman3}\cite{viswa1}
\ba
S[\phi_{0}]&=&
\frac{\Gamma
(1-\nu)}{\Gamma(\nu)}\nonumber\\ &\times&
\int\frac{d^{d}k}{(2\pi)^{d}}\;
\phi_{0}\left(\vec{k}\right)\phi_{0}\left(-\vec{k}\right)\;
\left(\frac{k}{2}\right)^{2\nu}\; ,\nonumber\\
\label{9012}
\ea
where $k=\mid\vec{k}\mid$, and performing the Legendre transform 
\cite{witten2}
\ba
{\tilde S}[\phi_{0},{\tilde \phi}_{0}] &=& S[\phi_{0}] \nonumber\\ 
&+&
\alpha\int\frac{d^{d}k}{(2\pi)^{d}}\;
\phi_{0}\left(\vec{k}\right){\tilde \phi}_{0}\left(-\vec{k}\right)\; ,
\label{9013}
\ea
where $\alpha$ is a coefficient, we arrive at 
\ba
{\tilde S}[{\tilde \phi}_{0}]
&=&-\frac{\alpha^{2}}{4}\;\frac{\Gamma   
(\nu)}{\Gamma(1-\nu)}\nonumber\\ &\times&
\int\frac{d^{d}k}{(2\pi)^{d}}\;
{\tilde
\phi}_{0}\left(\vec{k}\right){\tilde \phi}_{0}\left(-\vec{k}\right)\;
\left(\frac{k}{2}\right)^{-2\nu}\; .\nonumber\\
\label{9014}
\ea
It was verified in \cite{witten2} that the above generating 
functional gives rise 
to the missing conformal dimension $\Delta_{-}(\varrho)$.

However, there still remain some problems to be considered 
\cite{our2}\cite{our3}. One of them is that the Legendre transform 
Eq.(\ref{9013}) does not reproduce any of the constraints 
Eqs.(\ref{12000},\ref{9041}). And the another difficulty is that 
Eq.(\ref{9013}) does not work for $\nu =0$, due to the 
presence of a logarithmic term in the generating functional \cite{our2}.

In order to find a way out of these problems, we follow \cite{our3} and 
propose to consider a 
modified formulation of the scalar field theory on AdS and in the 
AdS/CFT correspondence. We begin by focusing on the definition of the 
energy. Note that the usual definition using the stress-energy tensor, 
as in \cite{freedman}\cite{freedman8}, is not sensitive to the addition 
of boundary terms to the action \cite{our3}, unlike what happens to the 
AdS/CFT prescription Eq.(\ref{9003}).\footnote{The relevance of the 
addition of boundary terms to the action in the AdS/CFT context 
was previously emphasized in the cases of the spinor field \cite{sfetsos} 
(were the boundary term proposed in \cite{sfetsos} was later computed 
using the Variational Principle \cite{frolov}\cite{henneaux}), the 
antisymmetric tensor field \cite{frolov2} and the Self-Dual model 
\cite{our}. Discussions using the Hamiltonian formalism can be found in 
\cite{frolov}} So, it seems 
natural to consider, 
at least in the AdS/CFT context, a new definition of energy which, 
unlike the usual one, is sensitive to the addition of boundary 
terms. Then, the natural choice is to define the energy as the 
`conserved' charge which is constructed out of the Noether current 
corresponding to time displacements in global 
coordinates. We shall refer to it as the `canonical energy', which is 
to be contrasted with the usual `metrical' one. Note that, unlike the 
metrical energy, the canonical energy is sensitive to the 
addition of boundary terms to the action, a property inherited from 
the Noether current. 

In order to perform the calculations, we consider the following action
\ba
I&=&-\frac{1}{2}\;\int\; d^{d+1} x \;\sqrt{g}\;
[ g^{\mu\nu}\partial_{\mu}\phi\;\partial_{\nu}\phi\nonumber\\ 
&&\qquad\qquad\qquad\qquad
+
\left( m^{2}+\;\varrho R\right)\phi^{2}]\nonumber\\ &+&\varrho\int\;
d^{d}x\;\sqrt{h}\; K\;\phi^{2}\; ,
\label{714}
\ea
where $h_{\mu\nu}$ is the induced metric at the boundary, and $K$ 
is the trace of the extrinsic curvature. Note that the 
above action differs from Eq.(\ref{9016}) only by a surface term 
which 
makes the action to be stationary under infinitesimal variations of the 
metric. Such term is, in fact, the natural extension of the usual 
Gibbons-Hawking term \cite{gibbons}. By performing an infinitesimal 
variation of the scalar 
field, it can also be verified that $I$ is stationary under 
Dirichlet boundary conditions which fix the scalar field at the 
boundary. Throughout this paper we will consider, with illustrative 
purposes, only Dirichlet boundary conditions, but in fact 
more complicated cases, such as Neumann and diverse mixed boundary 
conditions, can also be analyzed \cite{our2}\cite{our3}\cite{our4}.

It can be shown that, 
to perform a quantization which takes into account the canonical energy 
instead of the metrical one, gives rise to the following condition for 
the propagation of irregular modes in the bulk \cite{our3}
\beq
\varrho =\frac{1}{2d}\;\Delta_{-}(\varrho)\; .
\label{150}
\eeq
It has two possible solutions
\beq
\varrho^{\pm}=\frac{d-1}{8d}\;
\left[1\pm\sqrt{1+\left(\frac{4m}{d-1}\right)^{2}}\right]\; .
\label{151}
\eeq

The constraint Eq.(\ref{150}) is to be contrasted with the usual one 
Eq.(\ref{9041}). Note that, in the particular case $m=0$, 
$\varrho^{-}$ vanishes, whereas $\varrho^{+}$ reduces to the critical 
value $\frac{d-1}{4d}$ for which the stress-energy tensor becomes 
traceless. This means that, in particular, Weyl-invariant theories in 
the bulk allow for irregular modes to propagate.

Once we have considered modifications in the formulation in the bulk, 
the next step is to introduce a generalized AdS/CFT prescription which, 
in particular, is able to reproduce the constraint Eq.(\ref{12000}), 
together with the new one Eq.(\ref{150}). We will consider a modified 
prescription of the form \cite{our2}\cite{our3}
\ba
\exp\left( -I _{AdS}[f_{0}]\right)\qquad\qquad\qquad\quad \nonumber\\ 
\equiv 
\left<\exp\left(\int d^{d}x 
\;
{\cal O}(\vec{x}) \; f_{0}(\vec{x})\right)\right>,
\label{9044}
\ea
where the source $f_{0}$ which couples to the boundary conformal 
operator depends on the boundary conditions (in the particular 
Dirichlet case that we are considering in this paper, $f_{0}$ reduces 
to $\phi_{0}$). In addition, we will introduce a generalized Legendre 
transform prescription in which the Legendre transform is performed on 
the whole on-shell action, rather than only on the leading non-local term, 
as in Eq.(\ref{9013}). We do so because the operations of expanding the 
on-shell action in powers of the distance to the boundary and then 
selecting the leading non-local term, and of performing the Legendre 
transformation, are not commuting operations. Thus, the 
modified Legendre transform prescription schematically reads \cite{our3}
\ba
{\tilde I}_{AdS}[f_{0},{\tilde f}_{0}] &=& I_{AdS}[f_{0}] \nonumber\\
&-&
\int\frac{d^{d}k}{(2\pi)^{d}}\;
f_{0}\left(\vec{k}\right){\tilde f}_{0}\left(-\vec{k}\right)\; 
.\nonumber\\
\label{9045}
\ea
We emphasize that the above prescription contains information, in 
particular about Eqs.(\ref{12000},\ref{150}), that is being missed in 
the usual prescription Eq.(\ref{9013}). Once the modified Legendre 
transform has been performed, 
we should also include in our analysis the `conjugated' prescription of 
Eq.(\ref{9044}), which reads
\ba
\exp\left( -{\tilde I}_{AdS}[{\tilde 
f}_{0}]\right)\qquad\qquad\qquad\quad \nonumber\\ \equiv
\left<\exp\left(\int
d^{d}x \;
{\tilde {\cal O}}(\vec{x}) \; {\tilde 
f}_{0}(\vec{x})\right)\right>.
\label{9044'}
\ea
Following standard calculations in Euclidean Poincar\'e coordinates, it 
can be shown that the generalized prescription 
Eqs.(\ref{9044},\ref{9045},\ref{9044'}) gives rise, in a natural way, 
to the constraints Eqs.(\ref{12000},\ref{150}) (see \cite{our3} for 
details). In fact, in such situation the divergent local terms of the 
on-shell action cancel out, the addition of counterterms is not 
required, and the Legendre transform interpolates between different 
conformal dimensions, namely $\Delta_{+}(\varrho)$ 
and $\Delta_{-}(\varrho)$, as expected. 

Thus, we have considered two 
calculations which, at first sight, may seem very different of each 
other. One of them performs the quantization of the scalar field in 
global coordinates using the canonical energy instead of the usual 
metrical one, and the another one is concerned with calculations using 
the modified 
AdS/CFT and Legendre transform prescriptions. The fact that both 
formulations 
give rise, in a natural way, to precisely the same constraints 
Eqs.(\ref{12000},\ref{150}), could be considered as a strong evidence 
in 
support of our formalism. Further details on this subject can be 
found in \cite{our3}.

Another purpose of this paper is to show how this modified AdS/CFT 
prescription could be used to consistently describe 
double-trace perturbations at the boundary. It has 
recently been suggested \cite{berkooz1}\cite{berkooz3} that deforming
the boundary CFT by double-trace operators gives rise to a new   
perturbation expansion for string theory which is based on a non-local
worldsheet. This raises the question of how to perform explicit 
calculations in the AdS/CFT context including double-trace 
perturbations. The proposal in \cite{witten7}\cite{berkooz2} is that  
multi-trace operators can be incorporated by generalizing the usual 
Dirichlet prescription which is considered in the case of single-trace 
operators. In analyzing the case of a conformal operator of conformal 
dimension $\Delta=d/2$, it has been shown in \cite{witten7} that a 
generalized boundary condition gives rise to the correct 
renormalization formula for the coupling of the double-trace 
interaction. Further developments have also been introduced in 
\cite{muck}\cite{petkou}\cite{sever}\cite{barbon}\cite{berkooz8}\cite{gubser3}\cite{gubser4}\cite{comment}. 
The conjecture in \cite{witten7}\cite{berkooz2} that double-trace 
interactions change the boundary conditions of fields strongly suggests 
that the formalism developed in \cite{our2}\cite{our3} could be a 
natural frame to incorporate double-trace perturbations in the AdS/CFT 
correspondence. Such topic has been discussed in \cite{our4}. With 
illustrative purposes, we will here keep considering the situation of a 
Dirichlet boundary condition in the non-minimally coupled case. We will 
relate the coupling coefficient of the double-trace perturbation to the 
addition of specific boundary terms to the action. In this particular 
case, the natural extension of the Gibbons-Hawking term, as seen in 
Eq.(\ref{714}), will be generated. 

We introduce a perturbation at the boundary by a double-trace operator 
of the form
\beq   
W[{\cal O}]=\frac{\beta}{2}\; {\cal O}^{2}\; ,
\label{18}
\eeq
where $\beta$ is the coupling coefficient. Following the ideas in 
\cite{witten7}\cite{muck}, the perturbed generating 
functional can be written as \cite{our4}
\ba
I\left[f_{\epsilon}\right] &=& -\frac{1}{2}\int d^{d}x
\; d^{d}y\;\sqrt{h}\;f_{\epsilon}(\vec{x})
\;f_{\epsilon}(\vec{y})\nonumber\\ 
&\times&\int\frac{d^{d}k}{\left(
2\pi\right)^{d}}\;e^{-i\vec{k}\cdot\left(
\vec{x}-\vec{y}\right)}\;\frac{F(k\epsilon)}
{1+\beta(k\epsilon)F(k\epsilon)}\; ,\nonumber\\
\label{21}
\ea
where $\epsilon$ is a measure of the distance to the boundary which is 
considered to be small, and
\beq
F(k\epsilon)=\frac{d}{2} + \nu -
k\epsilon\;\frac{K_{\nu+1}(k\epsilon)}{K_{\nu}(k\epsilon)}\; .
\label{12}
\eeq
Here $K_{\nu}$ is the modified Bessel function. Following \cite{our4}, 
we are including all local and non-local terms in the action, and 
introducing a dependence of $\beta$ on $\epsilon$. This is needed for 
consistence. For $\beta =0$, Eq.(\ref{21}) reduces to the usual result. 
In addition to Eq.(\ref{21}), we should also include its Legendre 
transform in all calculations.

Now setting
\beq
\beta(k\epsilon)=\frac{2\varrho
d}{F(k\epsilon)}\;\frac{1}{F(k\epsilon)-2\varrho d}\; ,
\label{48}
\eeq
and introducing this into Eq.(\ref{21}), we reproduce precisely the 
same 
functional as the one computed in \cite{our3} when considering a 
Dirichlet boundary condition in the non-minimally coupled case. Then, 
the above double-trace perturbation generates the natural extension 
of 
the usual Gibbons-Hawking term, as seen in Eq.(\ref{714}). In 
particular, 
notice from the above equation that $\beta$ diverges precisely when 
the constraint Eq.(\ref{150}) is satisfied and irregular modes 
propagate in the bulk. This is the same situation for which the 
divergent local terms in the action cancel out, and the Legendre 
transform interpolates between different conformal dimensions. In 
particular, this result is consistent with the statement in 
\cite{witten7} that, as the coupling grows, the system approaches 
the condition that is suitable for quantization to get a field of 
dimension $\Delta_{-}$. For further 
details, and the analysis of other possible boundary conditions, see 
\cite{our4}. 

Thus, we have shown that the formulation in 
\cite{our3} consistently incorporates double-trace operators into the 
AdS/CFT correspondence, and we have also related the coupling 
coefficient of the double-trace perturbation to the constraints 
Eqs.(\ref{12000},\ref{150}) which arise when performing the modified 
quantization in the bulk, and to the addition of specific boundary 
terms to the action.


\begin{thebibliography}{9}
\bibitem{comment} Since the time this talk was given, additional 
references appeared \cite{witten5}\cite{troost}\cite{nolland}\cite{hoyos}.
\end{thebibliography}
\end{document}